\documentclass[journal,onecolumn]{IEEEtran}

\usepackage{times}

\usepackage{graphicx}
\usepackage{amsmath}
\usepackage{amsfonts}
\usepackage{subfigure}
\usepackage{verbatim}
\usepackage{algorithm}
\usepackage{algorithmic}

\newtheorem{definition}{Definition}

\begin{document}

\title{GCG: Mining Maximal Complete Graph Patterns from Large Spatial Data}
\author{
			\IEEEauthorblockN{Ghazi Al-Naymat}
			\IEEEauthorblockA{College of Computer Science and Information Technology\\
										  University of Dammam, KSA\\
										  ghalnaymat@ud.edu.sa}}


\maketitle

\begin{abstract}
Recent research on pattern discovery has progressed from mining frequent patterns and sequences to mining structured patterns, such as trees and graphs. Graphs as general data structure can model complex relations among data with wide applications in web exploration and social networks. However, the process of mining large graph patterns is a challenge due to the existence of large number of subgraphs. In this paper, we aim to mine only frequent complete  graph patterns. A graph $g$ in a database is complete if every pair of distinct vertices is connected by a unique edge. Grid Complete Graph (GCG) is a mining algorithm developed to explore interesting pruning techniques to extract maximal complete graphs from large spatial dataset existing in Sloan Digital Sky Survey (SDSS) data. Using a divide and conquer strategy, GCG shows high efficiency especially in the presence of large number of patterns. In this paper, we describe GCG that can mine not only simple co-location spatial patterns but also complex ones. To the best of our knowledge, this is the first algorithm used to exploit the extraction of maximal complete graphs in the process of mining complex co-location patterns in large spatial dataset. 
\end{abstract}

\section{Introduction} \label{sec:Introduction}

With the rapid invention of advanced technology, researchers have
been collecting large amounts of data on a continuous or periodic
basis in many fields. This data becomes the potential for
researchers to discover useful information and knowledge that has
not been seen before. In order to process this data and extract
useful information, the data needs to be organised in a suitable
format. Hence data preparation plays a very important role in the
data mining process.

The focus of this study is mainly on extracting interesting complex patterns from Sloan Digital Sky
Survey (SDSS) astronomy dataset~\cite{gray02}. The SDSS is the most
motivated astronomical survey project ever undertaken. The survey
maps in detail one-quarter of the entire sky, determining the
positions and absolute brightness of more than 100 million celestial
objects. The first official Data Release (DR1) of SDSS was in June
2003. Since then there have been many new releases including the
ninth major release (DR9) in August 2012 that provides images, imaging
catalogs, spectra, and redshift. Release DR9 contains more than 5TB of data, which includes measures of 500 million unique celestial
objects.

Availability of such large amount of useful data is an obvious
opportunity for application of data mining techniques to extract 
interesting information. However, while much research has been done
by the astronomical researchers, a feeble effort has been made to
apply data mining techniques on SDSS data. That is because the
SDSS data format is not suitable for mining purposes, that is \textit{the
main motivation of this paper}.

As mentioned in~\cite{chawla} spatial databases store spatial
attributes about objects, and hence, SDSS is a large spatial
dataset as it contains many attributes for each object. One of
the most significant problems in spatial data mining is to find
object types that frequently co-locate with each other in large
databases. Co-location means objects that are found in the
neighborhood of each other. The proposed approach mines co-location
patterns in SDSS data and uses these patterns to generate interesting
information about different types of galaxies. In this work only the galaxies existing in
SDSS is used. However, this approach could be generalised to be used with any other
celestial objects.

\begin{figure} [t]
        \begin{center}
                \includegraphics [height=2in,width=3.5in]{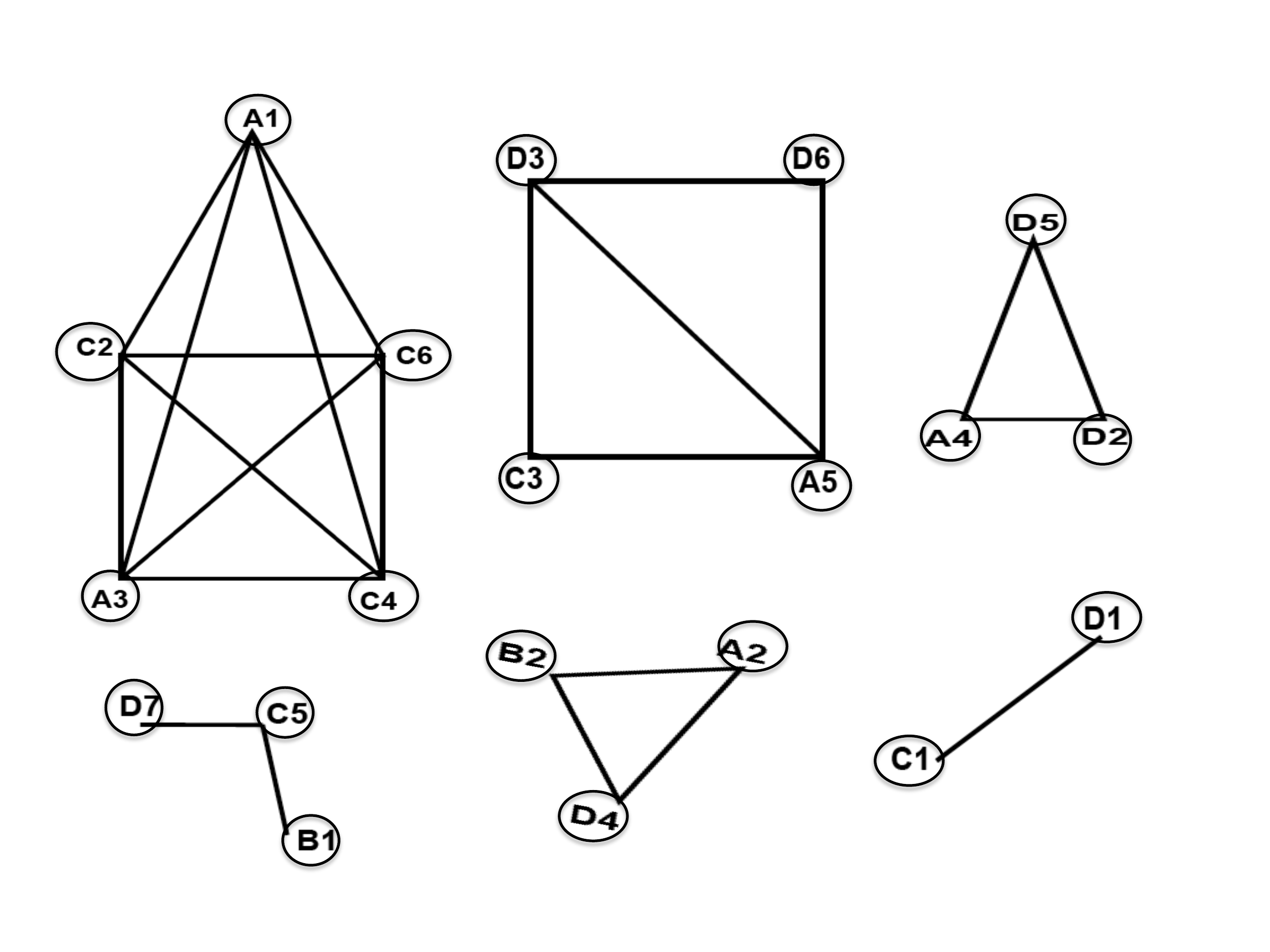}
                \caption{Graphs in a plane as examples of the co-location patterns}
                \label{Fig:colo}
         \end{center}
\end{figure}

The data preparation plays a vital rolein the mining process, hence it is done in two folds. First, extracting the galaxies in SDSS data and categorising
them into ``Early" and ``Late" type galaxies. Second, our proposed algorithm \emph{GCG} is utilised to generate co-location patterns (maximal complete graphs) from the data.  A complete graph is any set of
spatial objects such that all objects in the set co-locate. A
maximal complete graph is a complete graph which is not a subset of any other
complete graph. Fig.~\ref{Fig:colo} depicts some examples of spatial
co-locations, the line between the vertices (objects) indicates that
they are co-located. The second column of Table~\ref{Tab:CompleteGraphs} displays
the maximal complete graph patterns, which are presented in Fig.~\ref{Fig:colo}.

The full general problem of extracting
maximal complete graphs from a graph is known as NP-Hard. \emph{GCG}
efficiently extracts maximal complete graphs in a given spatial database
with capability to divide the space into grid structure based on a predefined
distance. A divide and conquer strategy is applied via a  grid structure to
reduce the search space.

\begin{table}[t]
\centering
 \caption{\label{Tab:CompleteGraphs} Maximal Complete Graph patterns}
   \begin{tabular}{cll}\hline\hline
     ID & Maximal Complete Graphs  & Transactions \\ \hline
        \hline
      1 & \{C1,D1\}             & \{C,D\}               \\ 
      2 & \{C5,D7\}             & \{C,D\}               \\ 
      3 & \{B1,C5\}             & \{B,C\}                \\ 
      4 & \{A4,D2,D5\}          & \{A,D+\}              \\
      5 & \{A5,C3,D3\}          & \{A,C,D\}             \\ 
      6 & \{A5,D3,D6\}          & \{A,D+\}             \\ 
      7 & \{A2,B2,D4\}          & \{A,B,D\}              \\ 
      8 & \{A1,A3,C2,C4,C6\}    & \{A+,C+\}             \\ \hline\hline
  \end{tabular}
\end{table}

In this paper we focus on using  maximal complete graphs to allow us to mine interesting \emph{complex spatial relationships} between the object types.  A \emph{complex spatial relationship} includes not only whether an object type, say $A$, is present in a (maximal) complete graph, but also:

\begin{itemize}
	\item Whether \emph{more than one} object of its type is present in the  \emph{maximal complete graph}. This is called a \emph{positive type} and is denoted by $A+$. 
	\item Whether objects of a particular type are not present in a \emph{maximal complete graph} -- that is, the absence of types. This is called a  \emph{negative type} and is denoted by $-A$.  
\end{itemize}

The inclusion of \emph{positive} and / or \emph{negative} \emph{types} makes a relationship \emph{complex}.
This allows us to mine patterns that say, for example, that $A$ occurs with multiple $B$'s but not with a $C$. That is, the presence of $A$ may imply the presence of multiple $B$'s and the absence of $C$. This is interesting in the astronomy domain. The last column of Table \ref{Tab:CompleteGraphs} shows examples of (maximal) complex relationships. 

Maximal complete graphs generated by \emph{GCG} can be
represented as transactions as given in Column 3 of
Table~\ref{Tab:CompleteGraphs}. These transactions can be used by ANY association rule mining technique.
Association rule mining techniques that proposed by~\cite{agrawal93,bavani05} can generate useful rules, which will be interpreted as relationships between objects.


We are not interested in \emph{maximal} complex patterns (relationships) in themselves, as they provide only local information (that is, about a maximal complete graph). We are however interested in \emph{sets} of object types (including complex types), that appear across the entire dataset (that is, amongst many maximal complete graphs). In other words, we are interested in mining \emph{interesting complex spatial relationships} (sets), where ``interesting'' is defined by a global measure. 
We use a variation of the $minPI$~\cite{agrawal94fast} measure to define interestingness.


\subsection{Problem Statement}\label{Problem}

{\bf Given the set of maximal complete graphs, find all interesting complex patterns  that occur amongst the set of maximal complete graphs. More specifically, find all \emph{sets} of object types, including \emph{positive} and \emph{negative} (that is, complex) types that are interesting as defined by their $Support$ being above a threshold.} 

This problem therefore becomes an \emph{itemset mining} task. In order to do this very quickly, we use \emph{interesting itemset mining algorithm}, that is GLIMIT~\cite{glimit06}.

Including negative types makes the problem much more difficult, as it is typical for spatial data to be sparse. This means that the absence of a type can be very common. 



\subsection{Contributions}

In this paper we make the following contributions:

\begin{itemize}

\item  An efficient algorithm called \emph{GCG} is proposed to generate all maximal complete graph patterns that exist in large spatial datasets.

 \item We introduce the concept of \emph{maximal complete graph}. We demonstrate how the use of \emph{maximal complete graphs} makes more sense than using complete graphs, and we showed that they allow the use of negative patterns.

    \item We show that complex and  interesting co-location patterns can be efficiently extracted from  huge and sparse spatial datasets.

\end{itemize}

The rest of the paper is organized as follows: Section~\ref{Sec:Method} gives further details of our approach. Section~\ref{Sec:Exps} contains our experiments and an analysis of the results. Section~\ref{Sec:RW} puts our contributions in context of related work, then we conclude in Section~\ref{Sec:Conc}.


\begin{figure*}[t]
    \centering
        \includegraphics[width=15cm]{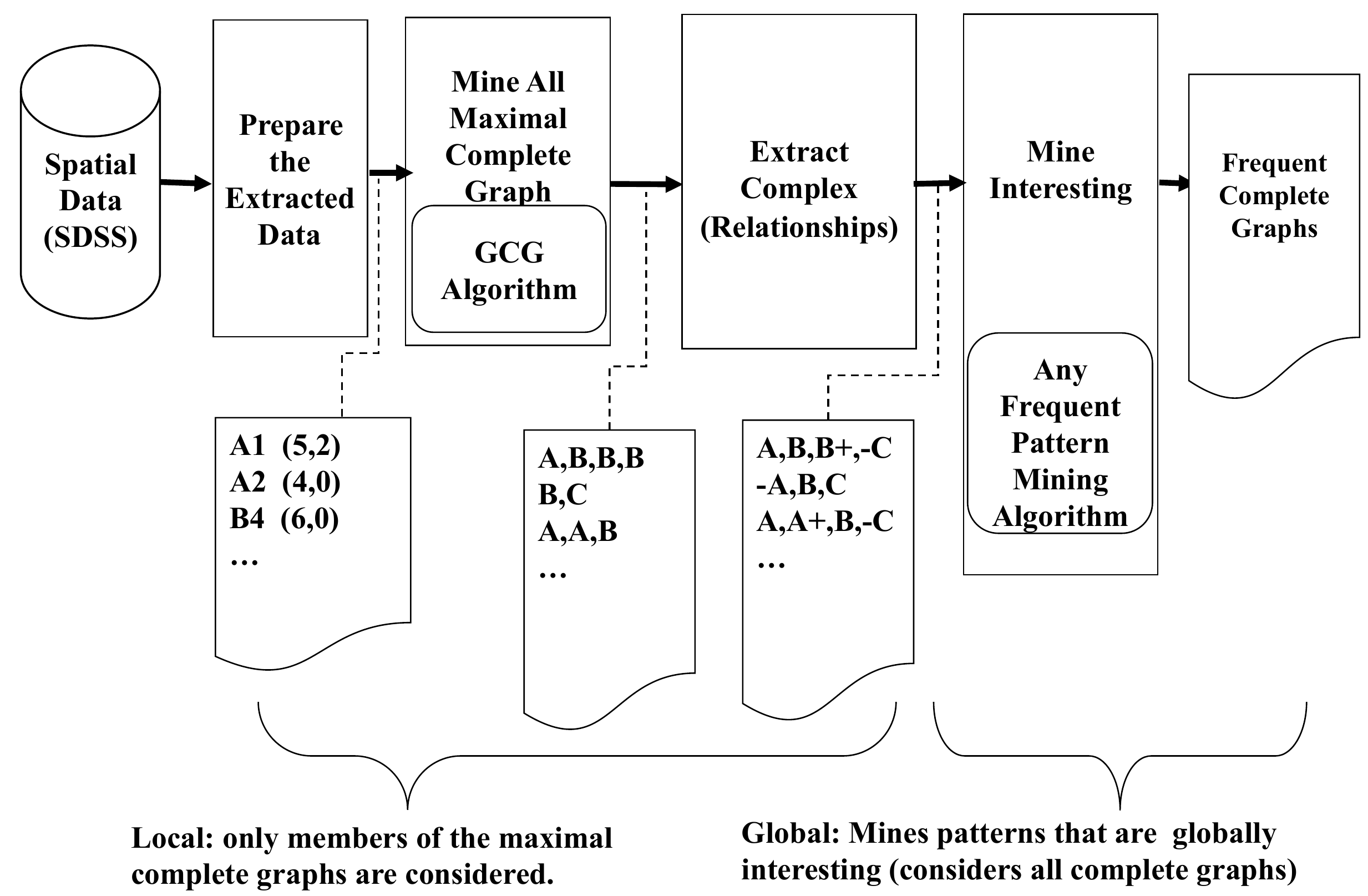}
        \caption{Framework showing the complete mining process. }
        \label{Fig:process}
\end{figure*}

\section{Maximal Complete Graph Mining} \label{Sec:Method}

Figure \ref{Fig:process} shows the overall process flow of our method. The follwoing subsections elaborate more about our approach.

\subsection{Data Extraction and Categorisation} 

Raw data needs most of the time to be prepared to suit data mining algorithms. This section illustrates the method of extracting the important attributes from the SDSS
database. These attributes used to categorise galaxy objects. A
view called \emph{SpecPhoto} which is derived from a table
called \emph{SpecPhotoAll} is used. The latter is a joined table between the
\emph{PhotoObjAll} and \emph{SpecObjAll} tables. In other words,
\emph{SpecPhoto} is view of joined \emph{Spectro} and \emph{PhotoObjects} that
have the clean spectra\footnote{http://www.sdss.org/}.

The concern was to extract only  the galaxy objects from the SDSS
using parameter (object type=0). The total number of galaxy-type
objects  stored in the SDSS catalog is more thant 507,594. However, to
ensure the accuracy for calculating the distance between objects
and the earth which leads to calculate the $X, Y,$ and $Z$
coordinates for each object, some parameters are used, such as
$zConf < 0.95$ (the rigid objects) and $zWarning =0$ (correct
RedShift). Therefore, the number of objects is reduced to (442,923).

SDSS release 9 provides a table called \emph{Neighbors}. This table
contains all objects that are located within 0.5 arcmins, this makes it 
not useful in this study because there is no
ability to choose any distance that would form the neighborhood
relation between objects. For example, in our experiments $(1, \cdots, 5)$ mega-parsec (distances) are used as the thresholds to check whether objects are close to each other or not. Table~\ref{Tab:schema} discloses the
extracted fields from the SDSS (DR9) that used during the preparation process.

\textbf{Data extraction:}
 The data was obtained from
SDSS (DR9)~\cite{SDSSData}. This data is extracted from the online
catalog services using several SQL statements and 
tools, which offered by the catalog. These tools are accessible
from the SDSS site\footnote{http://cas.sdss.org/dr5/en/tools/search/sql.asp}. 

\textbf{Data transformation:} The data obtained from the previous step is transformed  to identify the categories of the glaxies and to represent the data in the right format. 

 \begin{table}[t]
  \caption { The SDSS schema}\label{Tab:schema}
  \vspace{0.1in}
  \centering
    \begin{tabular}{|c|l|l|}\hline
        No  & Field name& Field description                 \\ \hline
                                                               \hline
        1.  &specObjID    & Unique ID                       \\ \hline
        2.  &z            & Final RedShift                  \\ \hline
        3.  &ra           & Right ascention                 \\ \hline
        4.  &dec          & Declination                     \\ \hline
        5.  &cx           & x of Normal unit vector         \\ \hline
        6.  &cy           & y of Normal unit vector         \\ \hline
        7.  &cz           & z of Normal unit vector         \\ \hline
        8.  &primTarget   & prime target categories         \\ \hline
        9.  &objType      & object type : Galaxy =0         \\ \hline
        10. &modelMag\_u  & Ultraviolet magniutde           \\ \hline
        11. &modelMag\_r  & Red Light magnitude             \\ \hline\hline
    \end{tabular}
 \end{table}

\textbf{New attributes creation:} With all necessary fields, this
step is to calculate the exact value of the $X, Y,$ and $Z$
coordinates which are not explicitly shown in the SDSS data.
First , the distance $D$ between objects and
the earth is calculated using Hubble's law and the value of $z$ for each
object as in Equation (~\ref{Hub} ) . Second, by considering
the unit vectors $cx, cy,$ and $cz$, and multiplying them by the
$D$, the value of $X, Y$ and $Z$ coordinates are calculated by
Equations ~\ref{XMPC}, ~\ref{YMPC}, and ~\ref{ZMPC}, respectively.

  \begin{equation}\label{Hub}
     D \approx \frac{c\times z}{H_{o}}
 \end{equation}

where $c$ is the speed of light, $z$ is the object RedShift, and
$H_{o}$ is Hubbles' constant. Currently the best estimate for this
constant is 71 $kms^{-1} Mpc^{-1}$~\cite{Hubble,Haynes}.

   \begin{equation}\label{XMPC}
     X = D \times cx
 \end{equation}

  \begin{equation}\label{YMPC}
     Y = D \times cy
 \end{equation}

  \begin{equation}\label{ZMPC}
     Z = D \times cz
 \end{equation}

\textbf{Galaxies Categorisation:} Different parameters were
used to categorise galaxy types. Based on the difference between
Ultraviolet $U$ and Red light magnitude $R$, galaxies are
categorised as either ``Early"' or ``Late"'. If the difference is greater
than or equal to $2.22$ the galaxy is ``Early"', otherwise it is ``Late"'. The value
of the \emph{r-band} \emph{Petrosian} magnitude
 indicates whether the galaxy is ``Main"' (close to the earth) or ``Luminous Red Galaxies" ($LRG$).
 That is by checking the value of \emph{r-band}. If \emph{r-band} $\leq17.77$, that indicates
 that the object is ``Main" galaxy otherwise it is ``LRG"~\cite{Martin02}.
 The four galaxy types that found are \textbf{Main-Late}, \textbf{Main-Early}, \textbf{LRG-Late}, and
\textbf{LRG-Early}.

\subsection{Basic Definitions and Concepts} \label{defs}

This section briefly defines the concepts that are used in this paper.
 
Consider a set of objects $O$ with fixed locations. Given an appropriate distance measure $d:O\times O\rightarrow \mathbb{R}$ we can define a graph $G$ as follows; let $O$ be the vertices and construct an edge between two objects $o_{1}\in O$ and $o_{2}\in O$ if $d(o_{1},o_{2})\le \tau$, where $\tau$ is a chosen distance. A \emph{co-location pattern} is a connected subgraph.

\begin{definition}[Complete Graph]
A Complete Graph $g\in O$ is any fully connected subgraph of $G$. That is, $d(o_{1},o_{2})\le \tau \,\, \forall \{o_{1},o_{2}\} \in g\times g$.
\end{definition}

For example, in Fig.~\ref{Fig:colo}, \{A4,D2,D5\} form a complete graph as each
object co-locates with each other. Similarly \{C5,D7\} form another complete graph.

As we have mentioned in Section~\ref{sec:Introduction} we use maximal complete graphs so that we can define and use complex patterns meaningfully and to avoid double counting. 

\begin{definition}[Maximal Complete Graph]
A maximal complete graph $G_{M}$ is a complete graph that is not a subset (sub-graph) of any other complete graph. 
\end{definition}

In Fig.~\ref{Fig:colo}, \{A4,D2,D5\} form a maximal complete graph as it is not a
 subset of another complete graph. However, \{A2,D4\} is not a maximal complete graph
 since it is a subset of the complete graph \{A2,B2,D4\}.

The mining of maximal complete graphs is done directly -- it does \emph{not} require mining all sub-complete graphs first.

 \begin{definition}[Complete Graph's Cardinality]
 It is the number of vertices in a complete graph, that is $|O|$. In other words, it is the size of the complete graph.  This value can be used to find the total number of edges $E$ (Equation~\ref{Equ:E}) that the complete graph can have.  The below equation shows that. 
  
  \begin{equation}\label{Equ:E}
    E = \dfrac{|O|(|O|-1)}{2},
 \end{equation}  
     where $|O|$ is the number of vertices.
 \end{definition}
 

\subsection{Mining Maximal Complete Graphs} \label{Sec:MiningMCG}

First, data preperation process starts a maximal complete graph mining algorithm to extract all \emph{maximal complete graphs}, and strips them of the object identifiers (producing raw maximal complete craphs as shown in Table~\ref{Tab:Relations}. 
One pass is then made over the raw maximal complete graphs in order to extract complex relationships. We describe this in Section~\ref{sec:extract-complex}. This produces complex maximal complete graphs. Each of these complex maximal
complete graphs is then considered as a \emph{transaction} $T$, and an \emph{interesting itemset mining algorithm}, using $minPI$ as the interestingness measure, is used to extract the interesting complex relationships. 

In \emph{itemset mining}, the dataset consists of a set of transactions $T$, where each transaction $t\in T$ is a subset of a set of \emph{items} $I$; that is, $t\subseteq I$. In our work, the set of complex maximal complete graphs (relationships) becomes the set of transactions $T$ (third column in Table~\ref{Tab:CompleteGraphs}). The items are the object types -- including the complex types such as $A+$ and $-A$. For example, if the object types are $\{A,B,C\}$, and each of these types is present and absent in at least one maximal complete graph, then $I=\{A,A+,-A,B,B+,-B\}$.
An interesting itemset mining algorithm mines $T$ for interesting itemsets. The support of an itemset $I'\subseteq I$ is the number of transactions containing the itemset: $support(I')=|\{t\in T:I'\subseteq t\}|$. So called \emph{frequent itemset mining} uses the support as the measure of interestingness. For reasons described in Section~\ref{sec:Introduction} we use \emph{minPI}~\cite{Shekhar04} which, under the mapping described above, is equivalent to 

\[
minPI(I')=\min_{i\in I'}\{support(I')/support(\{i\})\}
\]

Since $minPI$ is \emph{anti-monotonic},  we can easily prune the search space for interesting patterns. We adopted the method used in GLIMIT (\cite{glimit06, Verhein-Naymat}) to mine the interesting patterns from maximal complete graphs.  GLIMIT is a very fast and efficient itemset mining algorithm that has been shown to outperform Apriori like algorithms~\cite{agrawal94fast} and FP-Growth~\cite{HPY00FPgrowth}.

As shown in Fig.~\ref{Fig:process}, the complete graph generation and complex relationship extraction are local procedures, in the sense that they deal only with individual maximal complete graphs. In contrast,
the interesting pattern mining is global -- it finds patterns that occur across the entire space. 
Secondly, we consider subsets of maximal complete graphs only in the last step -- after the complex patterns have been extracted.

\begin{table}[tbh!]
 \caption { Example: Dataset of two dimensions.}\label{Tab:ex-sample}
 \centering
  \begin{tabular}{|c|c|c|}\hline
    Object type &X-Coordinate &Y-Coordinate   \\ \hline
                                                 \hline
      A1        &2.5          &4.5            \\ \hline
      A2        &6            &4              \\ \hline
      A3        &2            &9              \\ \hline
      B1        &1.5          &3.5            \\ \hline
      B2        &5            &3              \\ \hline
      B3        &5            &4              \\ \hline
      C1        &2.5          &3              \\ \hline
      C2        &6            &3              \\ \hline
      D1        &3            &9              \\ \hline
      D2        &7            &1.5            \\ \hline\hline
  \end{tabular}
\end{table}


\begin{algorithm}
 \caption{Grid Complete Graph algorithm.} \label{Alg:SM-GCG}
\begin{algorithmic}[1]
\small

 \REQUIRE Set of points $(P_{1},\cdots,P_{n})$,
Threshold $\tau$

\ENSURE A list of maximal Complete Graph patterns.
\COMMENT{\textbf{Generating grid structure.}}

\STATE $GridMap \leftarrow \phi$

\STATE $PointList \leftarrow \{P_{1},\cdots,P_{n}\}$

\FORALL{$P_{i}\in PointList$}
         \STATE Get the coordinates of each point $Pk_{x},Pk_{y},Pk_{z}$
         \STATE Generate the composite key (GridKey=($Pk_{x},Pk_{y},Pk_{z}$)).
         \IF{$GridKey \in GridMap$}
                \STATE $GridMap \leftarrow P_{i}$
         \ELSE
                \STATE $GridMap \leftarrow$ new GridKey
                \STATE $GridMap.GridKey \leftarrow P_{i}$
         \ENDIF
\ENDFOR

\COMMENT{\textbf{Obtaining the neighborhood lists.}}

\FORALL {$p_{i}\in GridMap$}
    \STATE $p_{i}.list \leftarrow \phi$
    \STATE $NeighborGrids \leftarrow$ (the 27 neighbor cells of $p_{i}$)
    \STATE $NeighborList \leftarrow \phi$
    \IF{$NeighborGrids_{i}.size()>1$}
        \FORALL{$p_{j} \in NeighborGrids_{j}$}
            \IF{EucDist $(p_{i},p_{j}) \leq \tau$}
                 \STATE $p_{i}.list \leftarrow p_{j}$ ($p_{i}, p_{j}$ are neighbors)
            \ENDIF
        \ENDFOR
    \ENDIF
    \STATE $NeighborList \leftarrow p_{i}.list $
\ENDFOR

 \COMMENT{\textbf{Pruning neighborhood list if at least one of its items violates the maximal Complete Graph definition.}}

\STATE $TempList \leftarrow \phi$

\STATE $MComplete GraphList \leftarrow \phi$

\FORALL {$Record_{i} \in NeighborList $}
    \STATE $RecordItems \leftarrow Record_{i}$
    \FORALL {$p_{i} \in RecordItems$}
        \FORALL {$p_{j} \in RecordItems $}
            \IF{$EucDist(p_{i},p_{j}) \leq \tau$}
                \STATE $Templist \leftarrow p_{j}$ ($p_{i}, p_{j}$ are neighbors)
            \ENDIF
        \ENDFOR
    \ENDFOR
    \STATE $MComplete GraphList \leftarrow Templist$
\ENDFOR

\end{algorithmic}
\end{algorithm}

\subsection{GCG algorithm} \label{sec:GCG}

\begin{figure}[th]
    \begin{center}
                \includegraphics [height=2in,width=0.7\textwidth]{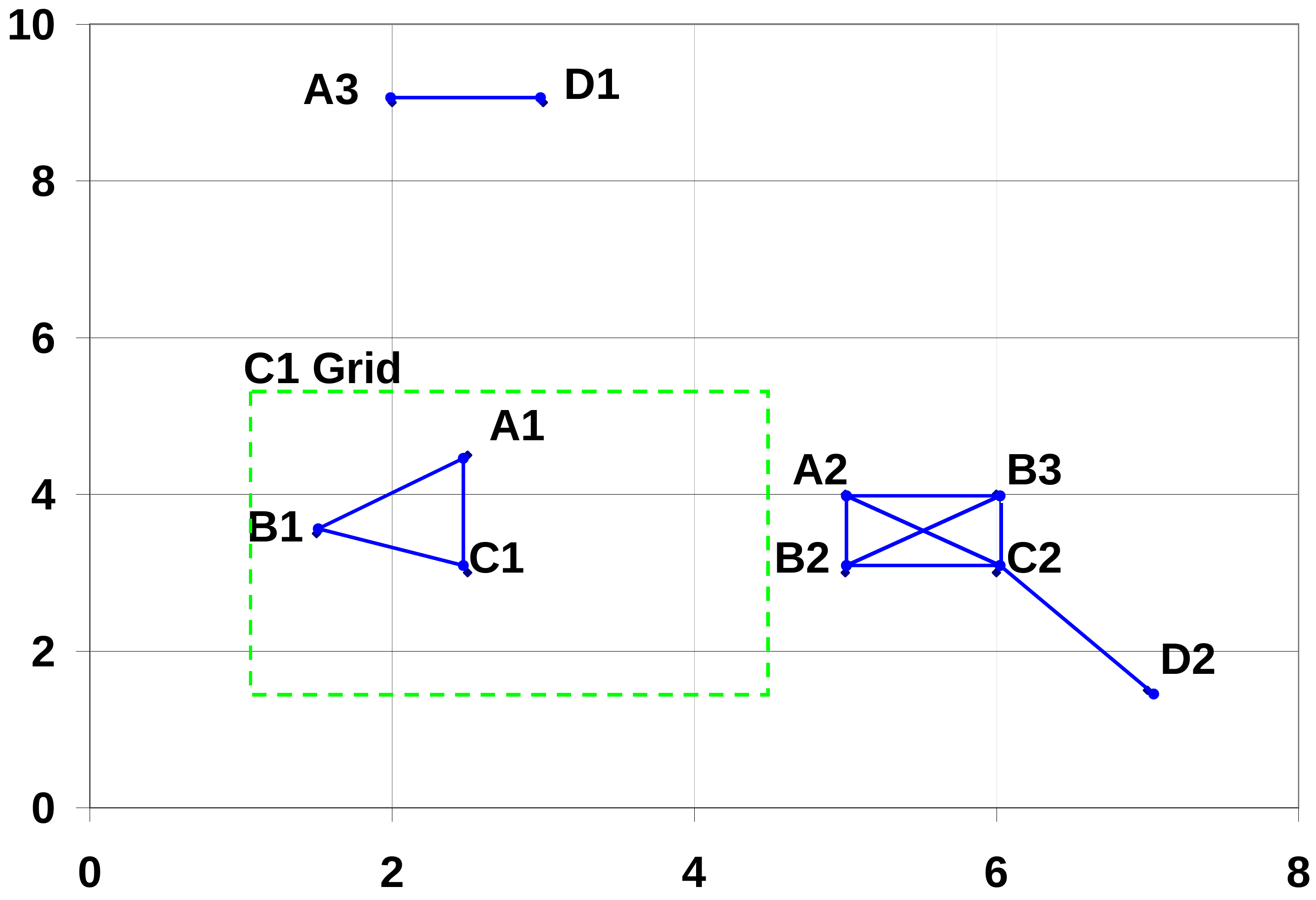}
                \caption{Spatial objects in a 2D (x and y coordinates) grid structure.}
                \label{Fig:GridStructure}
    \end{center}
\end{figure}

 Algorithm~\ref{Alg:SM-GCG} reveals the pseudocode of the \emph{GCG} algorithm. This section shows how the algorithm works through an example. By assuming that all objects are  spatial, we use Fig.~\ref{Fig:GridStructure} to depict some example items and their locations. These objects and their coordinates are given in Table~\ref{Tab:ex-sample}. It should be noted that SDSS is three dimensional dataset, but in the example two dimensions are used for the sake of simplicity.

\begin{table} [th!]
\centering
\caption{\label{Tab:Neighborhoodlists} Neighbor lists.} 
   \begin{tabular}{|l|l|}\hline 
     Lists &  Members \\ \hline 
     1      &  \{A1, B1, C1\} \\  \hline 
     2*    &  \{B1, A1, C1\}\\  \hline 
     3*    &  \{C1, A1, B1\}\\  \hline 
     4      &   \{D1, A3\}\\   \hline 
     5     &    \{A2, B2, C2, B3\}\\  \hline 
     6*    &  \{B2, A2, B3, C2\}\\  \hline 
      7**  &  \{C2, A2, B2, B3, D2\} \\  \hline 
      8    &  \{D2, C2\}\\   \hline 
      9*  &  \{A3, D1\}\\  \hline 
      10* &   \{B3, C2, B2, A2\} \\ \hline 
  \end{tabular} 
        
\end{table} 

\begin{table} [th!]
\centering
\caption{\label{Tab:Prunedlists} Neighbor lists after the pruning step. } 
   \begin{tabular}{|l|l|l|l|l|}\hline 
     Lists           &   1   		   & 4    		 &    5 &  8   \\ \hline 
     Members  &\{A1,B1,C1\}&\{D1,A3\}&\{A2,B2,C2,B3\}   &\{D2,C2\}\\ \hline
  \end{tabular} 
\end{table}

 Edges $E$ in each subgraph are formed by
 calculating the distance between adjacent objects. In other word, if the distance between them $\leq \tau$, the edge will be created. Each subgraph, in this context, forms a co-location pattern. Therefore, results of
this algorithm are patterns containing objects that are co-located. GCG algorithm (\ref{Alg:SM-GCG}) functionality is described as follows:

\begin{enumerate}
  \item \textbf{Lines 1 - 12}: Dividing the space into a grid structure and concurrently placing each point into its particular grid cell based on its coordinates (Fig.~\ref{Fig:GridStructure}). The size of the grid cell is $d\times d$, where $d = \tau$. The value of $\tau$ is given as one of the inputs for the GCG algorithm.

  \item \textbf{Lines 13 - 25}: Finding each object's neighborhood lists.
 This step is the most important one, and it is the most crucial step for the complexity issue. It uses the Euclidean distance technique to check the neighborhood relationship between objects. However, the number of checked spatial objects depends on the density of the grid and the content of the
 neighbor cells\footnote{Number of neighbor cells is 9 or 27 if the data is 2D or 3D, respectively.}. According to the example in Fig.~\ref{Fig:GridStructure}, also because the sample data contains 10 objects, a list for each object is created except
 for those objects that are located lonely. Our concern is to find co-location patterns that have number
 of members $\geq$ 2 (i.e. $|O| \geq 2$); because one object does not form any type of relationship. Consequently, no need to count objects that do not have connections (i.e. relationship) with at least one another object. However, in our
example all objects share relationships. For example, object \{A1\} has a relationship with \{B1,C1\} and object \{A2\} with \{B2, B3, C2\}. It can be seen that
  these objects share the same location, this means \{A1, B1, C1\} are co-located because the
  distance between them is $\leq \tau$. Table~\ref{Tab:Neighborhoodlists} shows some redundant lists -- marked by * -- (same objects in different order); this
  gives us the chance to prune the complete list without losing the objects as they present in another list. 
  
 \item \textbf{Lines 26 - 38}: Pruning any neighbor list that contains at least one object violating the co-location condition. For example, list 7 is  pruned because two of its members \{A2,D2\} are not close to each other as given in Table~\ref{Tab:Neighborhoodlists} (lists marked by **).
  
\end{enumerate}

  As a result of the previous steps, list of maximal complete graphs will be formed. For example, \{A1, B1, C1\} forms a maximal complete graph and so forth for lists (4, 5, 8) as shown in Table~\ref{Tab:Prunedlists}.

\subsection{GCG algorithm analysis}

This section discusses the GCG algorithm completeness, correctness, and complexity. 

\textbf{Completeness:} All objects in neighbor lists appear as
set or subset in maximal complete graph lists. After acquiring the entire neighbors for each point,
another check among these neighbors is done to assure that all points  are
neighbors to each other. Intuitively, doing that results to have
repeated neighbor lists. Therefore, this ensures finding
all maximal complete graphs in any given graph. 

\begin{figure}[thb!]
    \begin{center}
                \includegraphics [height=2in,width=3in]{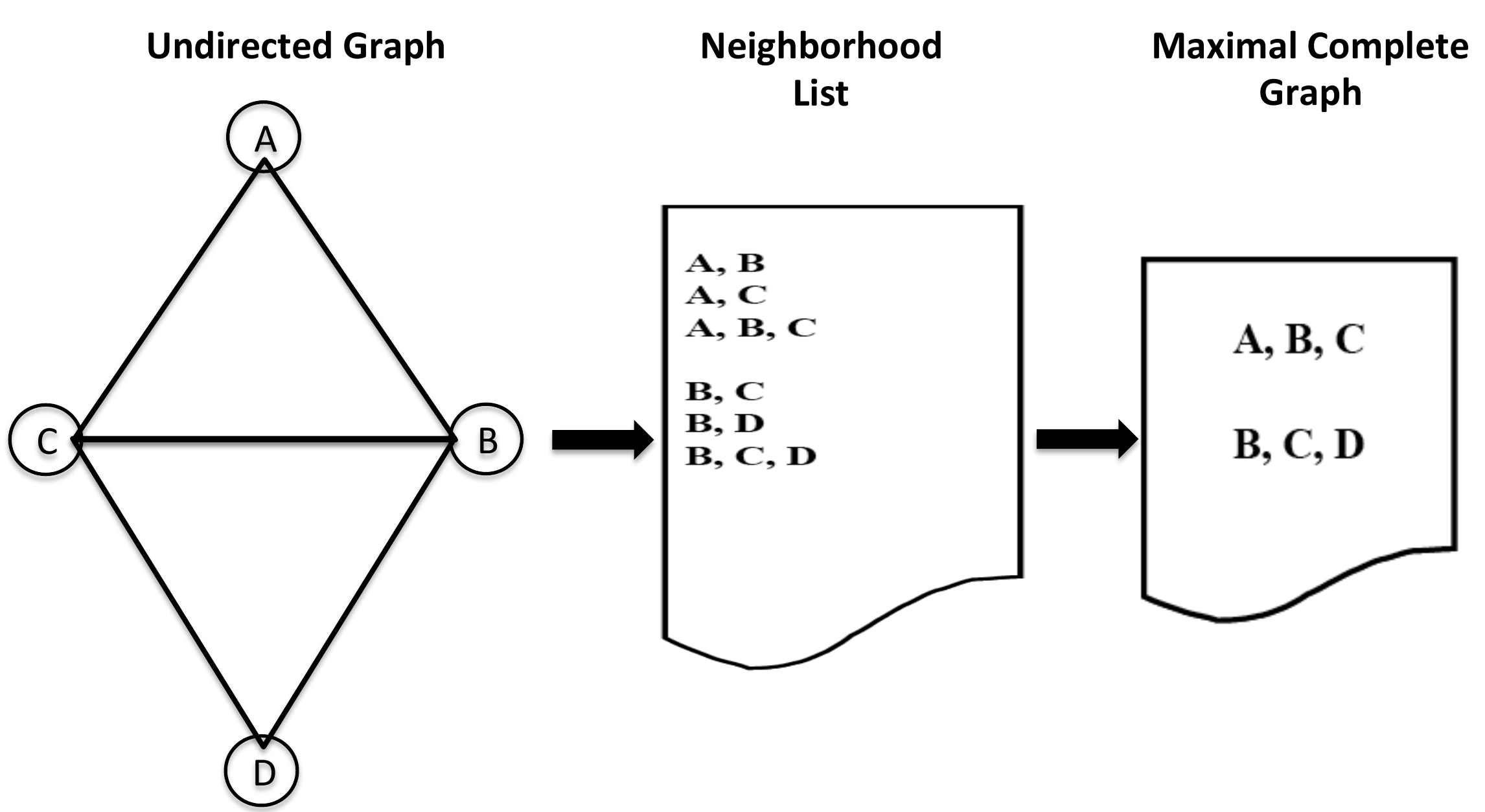}
                \caption{Example of two maximal complete graphs used to show the correctness of the proposed algorithm.}
                \label{Fig:CorrectnessExam}
    \end{center}
\end{figure}

\textbf{Correctness:} Every subset of a maximal complete graph appears
in the neighbors list. Thus, all maximal complete graphs that appear
in maximal complete graph's list will not be found as a subset in another
maximal complete graph. That is, the definition of maximal complete graph.
Fig.~\ref{Fig:CorrectnessExam} displays an undirect graph and the neighborhood
list and the existed maximal complete graph patterns. It is very clear that the pair $\{A,D\}$ does not appear in
the neighborhood list, because the distance between $d(A,D) > \tau$ (i.e. no edge between them). As a result, the pair $\{A,D\}$ will not be included in the maximal
complete graphs' list. In other words, any subset of any maximal complete graph
appears in the neighborhood list and it will not appear as an
independent  maximal complete graph. By this, the correctness
of the proposed algorithm is shown. 

\textbf{Complexity:} Assume there is $N$ points and
$c$ cells in a gird, and assume that all points are uniformly
distributed. Hence, on average there is $N/c$ points per cell. Also,
assume each cell has $l$ neighbors. Then to create the neighborhood
list of one point $l (N/c)$ points need to be examined to check if
they are within distance $\tau$. Since the total number of points is $N$,
thus the cost is $O (N^2l/c)$. And since $c >> l$, an assumption,
that this part of the algorithm is sub-quadratic, can be stated. Second, pruning neighborhood lists assuming that
on average the length of each neighborhood list is $k$. Then for
each neighborhood list, $k$ other lists have to be examined to check
if a point is in others neighborhood list or not. Therefore, for
each point,  $k$ other neighborhood lists are examined as well as
within each one, up to $k$ points will be checked. Consequently, the
cost is $O(N (k^2))$. Finally, the total cost is the cost to put the points in cell (O
(N)), the cost to create the neighborhood lists $O (N^2l/c) $, and
the cost to prune the lists $O(N (k^2))$. The total complexity of
the algorithm is $O(N(Nl/c+ k^2+ 1))$.


\begin{figure}[th]
    \centering
        \includegraphics[width=8cm]{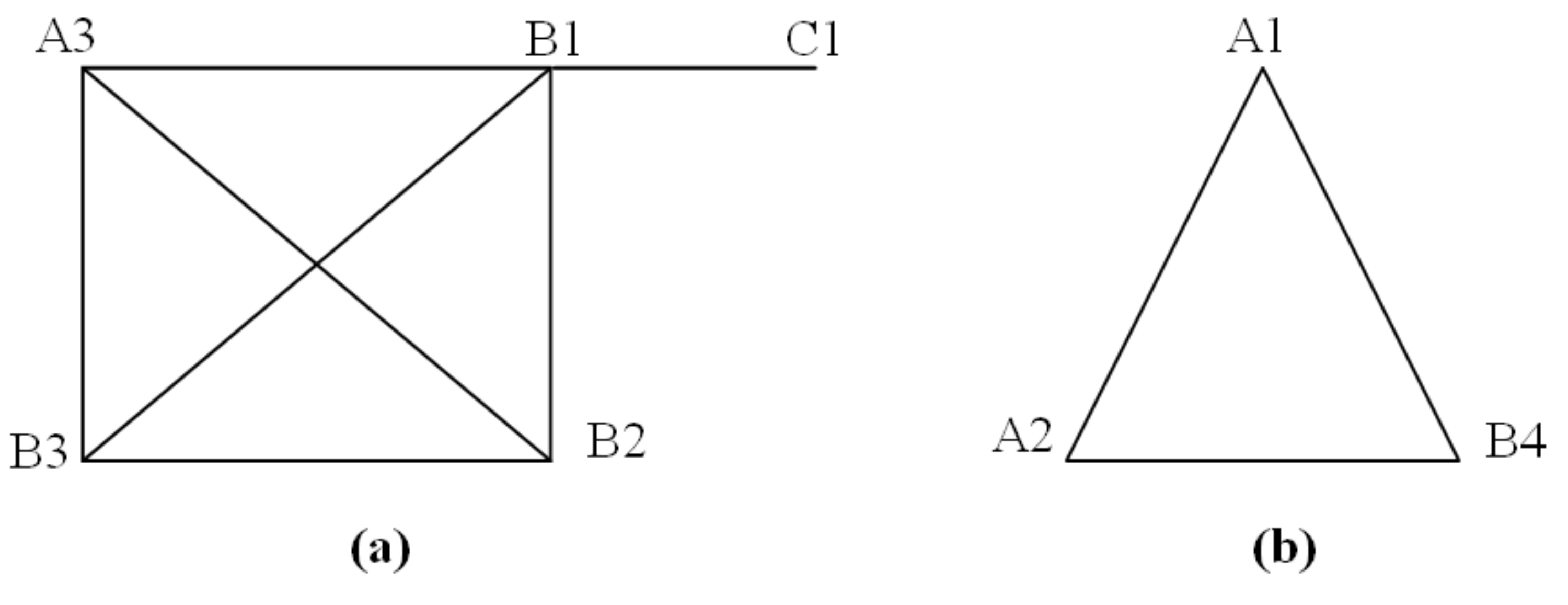}
        \caption{Complete graph example used in explaining the process of extracting the complex relationships.}
        \label{Fig:example}
\end{figure}

\begin{table} [thb!]
\centering 
\caption{\label{Tab:Relations} Representing maximal complete graphs of Fig.~\ref{Fig:example} as complex relationships} 
   \begin{tabular}{|c|l|l|l|}\hline 
     ID & Maximal    					   & Raw Maximal           &  Complex       \\ 
         &       Complete Graphs       & Complete Graphs    &  Relationships      \\ \hline 
        \hline 
      1 & \{A3, B1, B2, B3\}                  & \{A, B, B, B\}                & \{A, B, B+, -C\} \\ \hline 
      2 & \{B1, C1\}                             & \{B, C\}                        &  \{-A, B, C\}     \\ \hline 
      3 & \{A1, A2, B\}                        & \{A, A, B\}                    &\{A, A+, B, -C\} \\ \hline 
  \end{tabular} 
\end{table} 

\subsection{Extracting Complex Relationships} \label{sec:extract-complex}
A relationship is called complex if it consists of \emph{complex types} as defined in Section \ref{sec:Introduction}. 

Extracting a complex relationship $R$ from a maximal complete graph $G_{M}$ is straightforward -- we simply use the following rules for every type $t$:

\begin{enumerate}
	\item First, remove the object identifiers. This produces a ``raw'' maximal complete graph $G_{M}$.

	\item If $G_{M}$ contains an object with type $t$, $R = R\cup t$.

	\item If $G_{M}$ contains more than one object of type $t$, $R = R \cup t+$.

	\item If $G_{M}$ does not contain an object of type $t$, $R = R \cup -t$.

\end{enumerate}

Note that if $R$ includes a positive type $A+$, it will also \emph{always} include the basic type $A$. This is necessary to that maximal complete graphs that contain $A+$ will also be counted as containing $A$ when we mine for interesting patterns. 
 
Recall that the negative type only makes sense if we use \emph{maximal complete graphs}. The last column of Table~\ref{Tab:Relations} shows the result of applying all four rules.

 \begin{figure}[b!]
    \begin{center}
      \includegraphics [height=2in,width=0.7\textwidth]{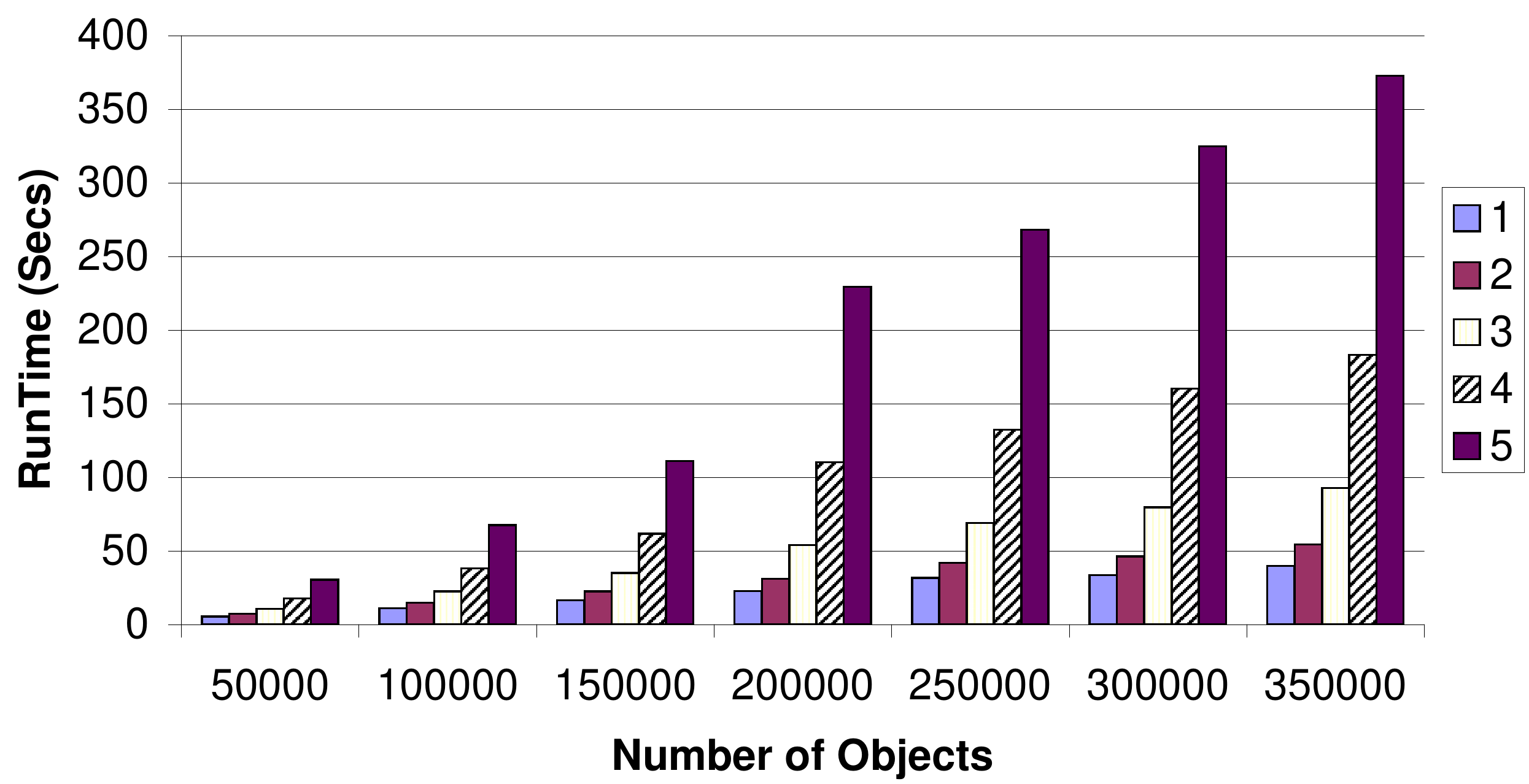}
      \caption{GCG's runtime using 5 different distances.}
      \label{Fig:GRun}
    \end{center}
\end{figure}

\begin{figure*}[thb!]
   \subfigure[Number of ``Main-Late" galaxies in complete graphs] {\includegraphics [height=3in,width=3.5in]{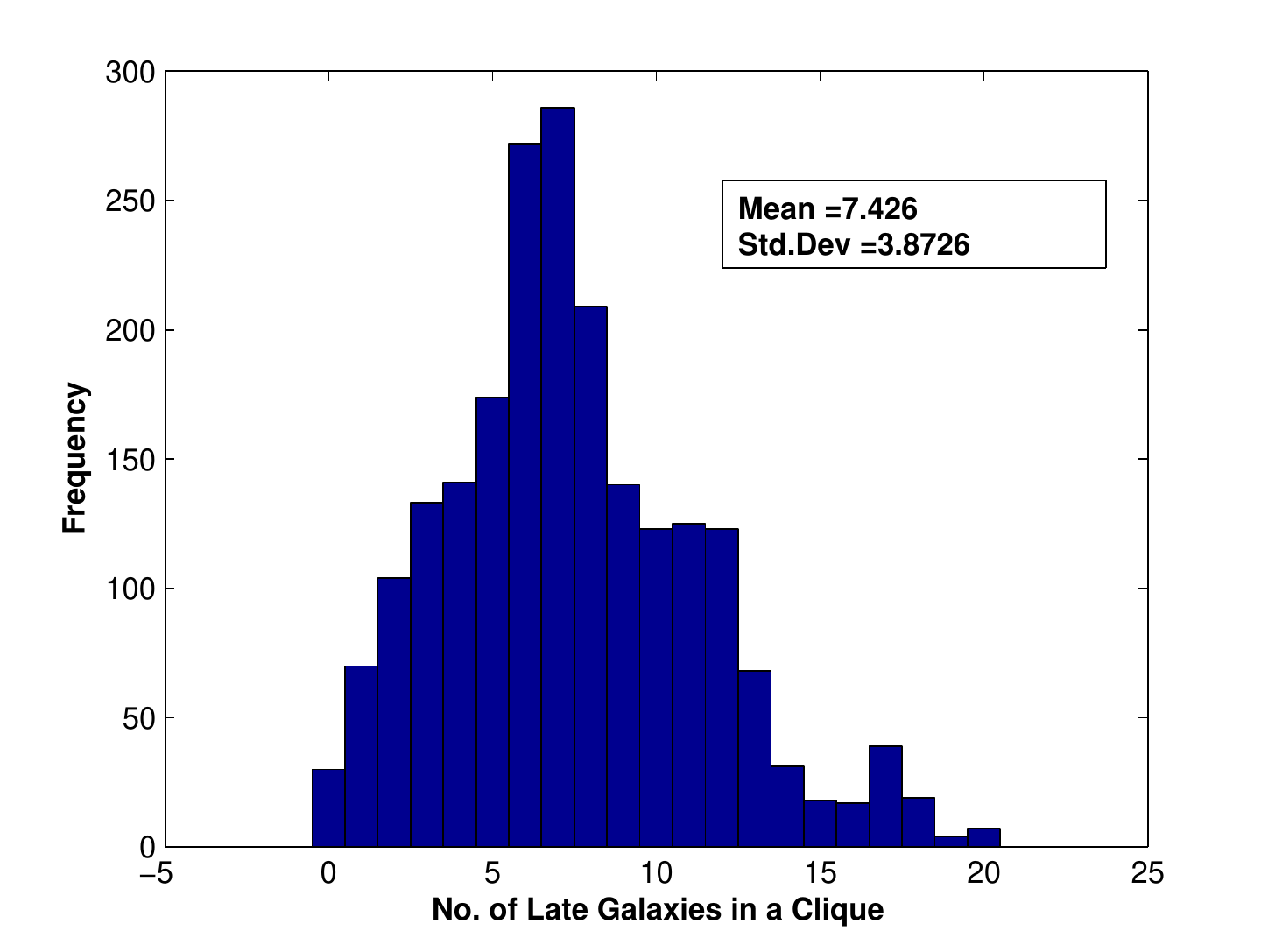}}
  \subfigure[Number of ``Main-Early" galaxies in complete graphs]{\includegraphics [height=3in,width=3.5in]{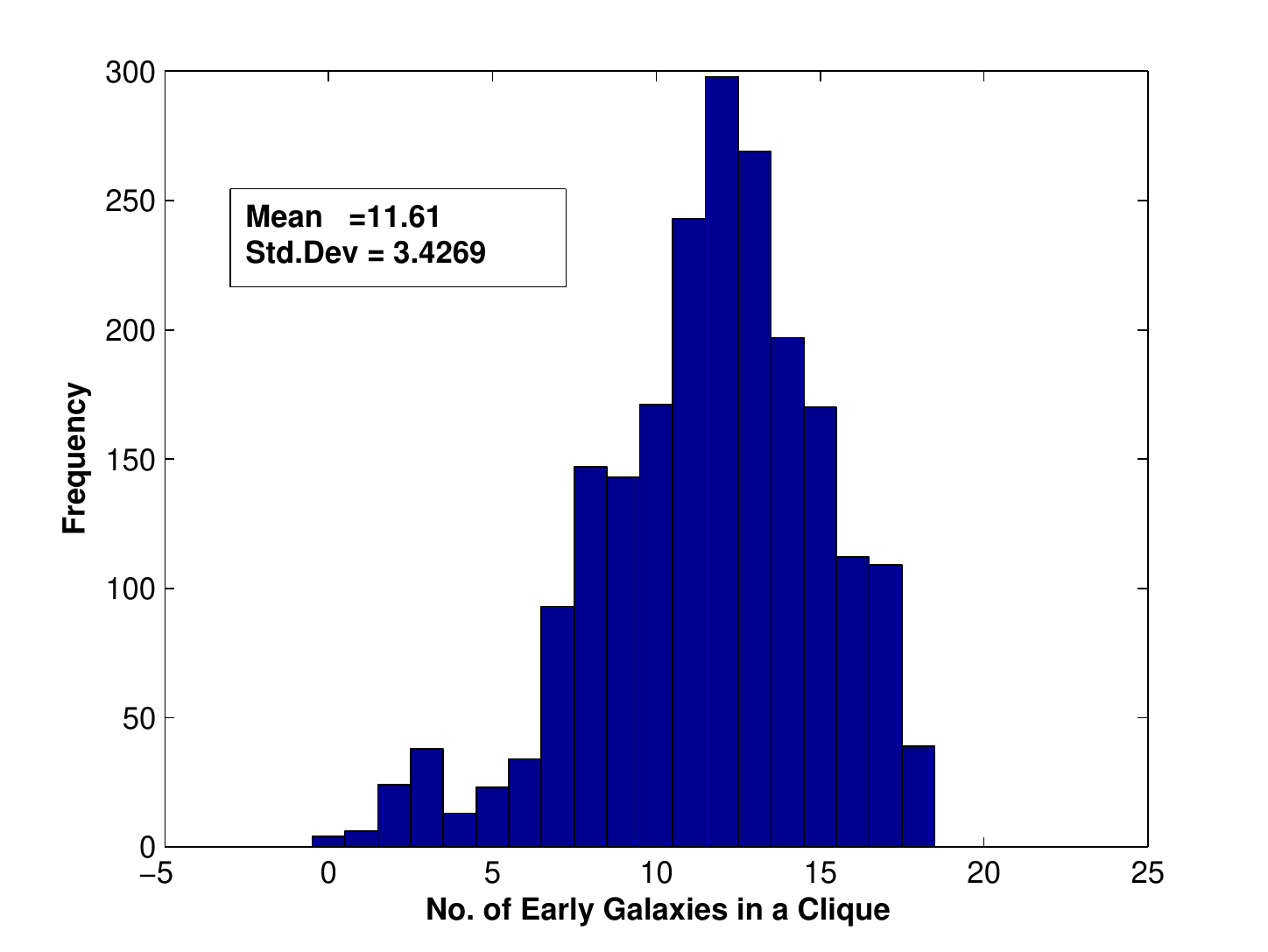}}
    \caption{The existence of galaxies in the universe.}
    \label{Fig:LargeGalaxies}
\end{figure*}

\section{Experiments and Results Discussion}\label{Sec:Exps}

 Experiments are carried out to confirm the achieved results when using the proposed algorithm on the SDSS data. All experiments were carried out on a Mac OS X 10.7 operated laptop (2.53 GHz) Intel Core Duo processor and 4 GB main memory. The data structures and algorithm were implemented in Java and compiled with the GNU compiler.


\subsection{Scalability of GCG algorithm}

Fig.~\ref{Fig:GRun} demonstrates the runtime of the \emph{GCG}
algorithm with various numbers of objects (galaxies) and distances. It illustrates that the runtime increases slightly as the number of objects and distance increase. The distance is increased by 1 Mpc every time, whereas the number of
objects is increased by 50K objects. The maximum number of records was 350000.  To explain further, when the distance increases the grid size increases. Also by increasing number of objects at the same
time, it allows more objects to appear in the same gird's cell or in
the neighbor grid areas. Therefore, the two factors
(distance, number of objects) affect the runtime of the \emph{GCG} algorithm.


\subsection{Galaxy types in large complete graphs}
We applied the \emph{GCG} algorithm on the ``Main" galaxies
extracted from SDSS to generate maximal complete graphs with neighborhood
distance (4 Mpc). We selected the complete graphs with the largest
cardinality ($|O|=22$). Fig.~\ref{Fig:LargeGalaxies} shows the distribution
of ``Early" and ``Late" type galaxies in the reported complete graphs. These
results show that large complete graphs consist of more ``Early" type galaxies
(Elliptic) than ``Late" type galaxies (Spiral). This conforms to the patterns given by~\cite{gray02} that say ``Early" type galaxies tend to stay away from ``Late" type galaxies.


\subsection{Complete Graphs Cardinalities}
\begin{figure} [t]
  \begin{center}
  \includegraphics[height=2in,width=0.7\textwidth]{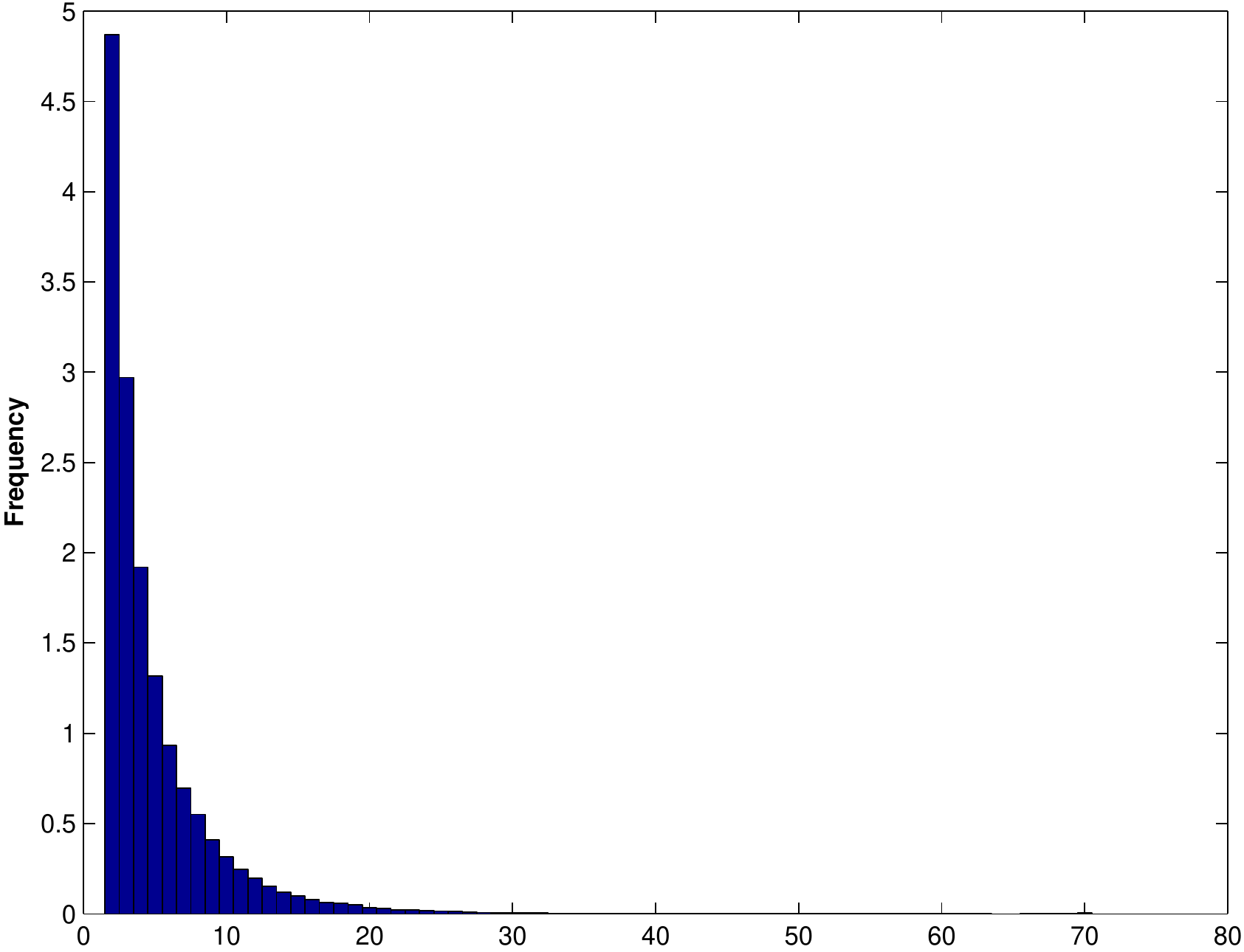}
  \caption{Complete Graphs cardinalities for Main galaxies using threshold = 4 Mpc. Frequency X $10 ^{4}$}\label{FIg:CompleteGraphsSize}
  \end{center}
\end{figure}

Figure~\ref{FIg:CompleteGraphsSize} shows the complete graphs cardinalities in ``Main" galaxies. It shows that complete graphs with cardinality between 2 and 5, small complete graphs, are more frequent than large complete graphs.

%
%

\section{Related Work}\label{Sec:RW}

Huang et al.~\cite{Huang03} defined the co-location pattern as the presence of a spatial feature in the neighborhood of instances of other spatial features. They developed an algorithm for mining valid rules in spatial databases using an Apriori based approach. Their algorithm does not separate the co-location mining and interesting pattern mining steps  like our approach does. 
Also, they did not consider complex relationships or patterns. 

Monroe et al.~\cite{Rob} used cliques as a co-location pattern (subgraphs), but in our research we used complete graphs instead. Similar to our approach, they separated the clique mining  from the pattern mining stages. However, they did not use maximal complete graph. They treated each clique as a transaction and  used an Apriori based technique for mining association rules. Since they used cliques (rather than maximal complete graphs) as their transactions, the counting of pattern instances is very different. They considered complex relationships within the pattern mining stage.  However, their definition of negative patterns is very different -- they used infrequent types while we base our definition on the concept of absence in \emph{maximal complete graphs}. They also used a different measure, namely, maxPI.

Arunasalam et al.~\cite{bavani05} used a similar approach to~\cite{Rob}. They proposed an algorithm called NP\_maxPI which also used the MaxPI measure. The proposed algorithm prunes the candidate itemsets using a property of maxPI.  They also used an Apriori based technique to mine complex patterns. A primary goal of their work was to mine patterns which have low support and high confidence. As with the work of~\cite{Rob}, they did not use maximal complete graphs. 

Zhang et al. \cite{Xin04} enhanced the algorithm proposed in~\cite{Huang03} and used it to mine special types of co-location relationships in addition to cliques, namely; the \emph{spatial star}, and \emph{generic} patterns. This means they didn't use maximal complete graphs. 

Most of the previous research and to the best of our knowledge, previous work has used Apriori type algorithms for mining interesting co-location patterns.  However, we embedded GLIMIT~\cite{glimit06} as the underlying pattern mining algorithm as already discussed in Section~\ref{Sec:MiningMCG}. To the best of our knowledge, no previous work has used the concept of \emph{maximal complete graph to mine comoplex co-location patterns in large spatial data}.


\section{Conclusion}\label{Sec:Conc}

In this paper, we presented a framework, which incorporates our proposed  algorithm \emph{GCG}   to mine complex co-location patterns exist in large spatial dataset (SDSS). Most of the previous research conducted in this area used Apriori type algorithms to mine only normal co-location patterns. However, we showed the importance of using complex co-location patterns, which are extracted from maximal complete graphs. We also presented how our proposed algorithms strips efficiently all  maximal complete graphs in large spatial dataset (SDSS) using divide and conquer strategy. We have shown that the idea of mining maximal complete graphs is very important in our work since complex patterns only makes sense when using maximal complete graphs. Future work would be to extend this framework to extract interesting relationships using different types of spatial objects in the astronomy domain. 


\bibliographystyle{IEEEtran}
\bibliography{latex8}
\end{document}